\documentclass[preprint,final,5p,times,twocolumn]{elsarticle}

\usepackage{amssymb}

\usepackage{graphicx}
\usepackage{multirow}
\usepackage{color}
\usepackage{amsthm}
\usepackage{amsmath}
\usepackage{comment}
\usepackage{threeparttable}
\usepackage[none]{hyphenat}

\journal{Physics Letters B}

\begin{document}

\begin{frontmatter}



\title{Fine structure in the $\alpha$ decay of $^{223}$U}

\author[IMP,UCAS,LZU]{M.D. Sun}
\author[IMP]{Z. Liu}
\author[IMP]{T.H. Huang}
\author[IMP]{W.Q. Zhang}
\author[UY,JAEA]{A.N. Andreyev}
\author[IMP]{B. Ding}
\author[IMP]{J.G. Wang}
\author[IMP,UCAS]{X.Y. Liu}
\author[IMP,UCAS]{H.Y. Lu}
\author[IMP,UCAS]{D.S. Hou}
\author[IMP]{Z.G. Gan}
\author[IMP]{L. Ma}
\author[IMP]{H.B. Yang}
\author[IMP]{Z.Y. Zhang}
\author[IMP]{L. Yu}
\author[IMP,UCAS]{J. Jiang}
\author[IMP,UCAS]{K.L. Wang}
\author[IMP]{Y.S. Wang}
\author[IMP]{M.L. Liu}
\author[PKU]{Z.H. Li}
\author[PKU]{J. Li}
\author[PKU]{X. Wang}
\author[IMP]{A.H. Feng}
\author[CIAE]{C.J. Lin}
\author[CIAE]{L.J. Sun}
\author[CIAE]{N.R. Ma}
\author[IMP]{W. Zuo}
\author[IMP]{H.S. Xu}
\author[IMP]{X.H. Zhou}
\author[IMP]{G.Q. Xiao}
\author[KTH]{C. Qi}
\author[BNU,BRC]{F.S. Zhang}

\address[IMP]{CAS Key Laboratory of High Precision Nuclear Spectroscopy, Institute of Modern Physics, Chinese Academy of Sciences, Lanzhou 730000, China}
\address[UCAS]{University of Chinese Academy of Sciences, Beijing 100049, China}
\address[LZU]{Lanzhou University, Lanzhou 730000, China}
\address[UY]{Department of Physics, University of York, York, YO10 5DD, United Kingdom}
\address[JAEA]{Advanced Science Research Centre (ASRC), Japan Atomic Energy Agency (JAEA), Tokai-mura, Japan}
\address[PKU]{State Key Laboratory of Nuclear Physics and Technology, School of Physics, Peking University, Beijing 100871, China}
\address[CIAE]{China Institute of Atomic Energy, P.O. Box 275(10), Beijing 102413, China}
\address[KTH]{KTH Royal Institute of Technology, Albanova University Center, SE-10691, Stockholm, Sweden}
\address[BNU]{Key Laboratory of Beam Technology and Material Modification of Ministry of Education, College of Nuclear Science and Technology, Beijing Normal University, Beijing 100875, China}
\address[BRC]{Beijing Radiation Center, Beijing 100875, China}

\begin{abstract}
Fine structure in the $\alpha$ decay of $^{223}$U was observed in the fusion-evaporation reaction $^{187}$Re($^{40}$Ar, p3n) by using fast digital pulse processing technique. Two $\alpha$-decay branches of $^{223}$U feeding the ground state and 244 keV excited state of $^{219}$Th were identified by establishing the decay chain $^{223}$U $\xrightarrow{\alpha_{1}}$ $^{219}$Th $\xrightarrow{\alpha_{2}}$ $^{215}$Ra $\xrightarrow{\alpha_{3}}$ $^{211}$Rn. The $\alpha$-particle energy for the ground-state to ground-state transition of $^{223}$U was determined to be 8993(17) keV, 213 keV higher than the previous value, the half-life was updated to be 62$^{+14}_{-10}$ $\mu$s. Evolution of nuclear structure for $N$ = 131 even-$Z$ isotones from Po to U was discussed in the frameworks of nuclear mass and reduced $\alpha$-decay width, a weakening octupole deformation in the ground state of $^{223}$U relative to its lighter isotones $^{219}$Ra and $^{221}$Th was suggested.
\end{abstract}


\end{frontmatter}

\section{\label{sec:level1}Introduction}

Light actinide nuclei with neutron number $N$ $\sim$ 134 are susceptible to effects of octupole deformation~\cite{Butler1996,Butler2016}. Alpha decay is long known as one of the tools to search for the effects of octupole correlations in nuclei. In particular, the first experimental evidence for octupole deformation in the actinides came from fine structure $\alpha$ decays feeding low-lying 1$^-$ and 3$^-$ states in the $A$ = 220-224 isotopes of Ra~\cite{Asaro1953} and Rn~\cite{Poynter1989}. Static or dynamic octupole deformations were expected to exist in light actinides with mass number $A$ between 219 and 229~\cite{Sheline2002}. The strongest octupole coupling for actinides was supposed to occur at $Z$ $\sim$ 88 and $N$ $\sim$ 134~\cite{Butler2016}, heavier than the doubly magic nucleus $^{208}$Pb in the chart of the nuclides. When going towards $^{222}$Ra from $^{208}$Pb, nuclear ground-state shape undergoes three phases: spherical, quadrupole-octupole and quadrupole deformed shapes. The $N$ = 131 isotones are situated in a transitional region between $N$ = 126 spherical and $N$ = 134 octupole deformed nuclei.

The $N$ = 131 isotones of $Z$ = 84 and 86, i.e., $^{215}$Po and $^{217}$Rn, have nearly spherical ground states with J$^{\pi}$ = 9/2$^{+}$, corresponding to the odd neutron in the $\nu g_{9/2}$ orbital. In the $\alpha$ decay of $^{215}$Po and $^{217}$Rn, ground-state to ground-state transition comprises nearly 100\% of the total $\alpha$-decay intensity, reflecting the remarkable similarity between the parent and daughter configurations.

In heavier $N$ = 131 isotones $^{219}$Ra ($Z$ = 88) and $^{221}$Th ($Z$ = 90), experimentally fine structure in the $\alpha$ decays were observed. Onset of quadrupole-octupole correlations was predicted in the framework of the reflection-asymmetric rotor model~\cite{Leander1988}. Moreover, dynamic octupole correlations were indicated in the $\gamma$ spectroscopy in $^{219}$Ra~\cite{Hensley2017} and $^{221}$Th~\cite{Reviol2014}. 

Nonzero octupole deformation was predicted in the ground state of $^{223}$U ($Z$ = 92) by macroscopic-microscopic model~\cite{Moller2016}, while the octupole-deformed minimum is not the lowest in energy according to covariant density functional theory~\cite{Agbemava2016}. Experimental study of $^{223}$U will shed new light on the shape evolution of $N$ = 131 and 129 ($\alpha$-decay daughter) isotones.

Isotope $^{223}$U was first identified by the Dubna group at the electrostatic separator VASSILISSA in the reaction of $^{208}$Pb($^{20}$Ne, 5n)~\cite{Andreyev1991}. Based on $\alpha$-$\alpha$ correlation technique and by searching for the decay chains of $^{223}$U $\xrightarrow{\alpha}$ $^{219}$Th $\xrightarrow{\alpha}$ $^{215}$Ra $\xrightarrow{\alpha}$ $^{211}$Rn, an $\alpha$-particle energy of 8780(40) keV and a half-life of 18$^{+10}_{-5}$ $\mu$s were reported for $^{223}$U. The decay energy was extracted by subtracting the known $\alpha$-particle energy of $^{219}$Th from the sum energy (full pile-up) of $\alpha$ decays of $^{223}$U and its fast-decaying daughter $^{219}$Th ($T_{1/2}$ = 1.05 $\mu$s) measured by using traditional analog electronics. 

In the present work, we report a new experimental study of $^{223}$U using a fast digital pulse processing technique which allows to clearly separate the fast sub-microsecond decays, which was the main difficulty in the previous studies. Thanks to the use of this method, fine structure in the $\alpha$ decay of $^{223}$U was observed for the first time. Structural evolution in $N$ = 129 and 131 isotones will be discussed in the framework of the reduced $\alpha$-decay width.

\section{\label{sec:level2}Experiment and results}

The experiment was performed at the gas-filled recoil separator SHANS \cite{Zhang2013} at IMP (Lanzhou, China). Isotope $^{223}$U was produced in the fusion-evaporation reaction $^{187}$Re($^{40}$Ar, p3n) at a beam energy of 188 MeV with an average beam intensity of 320 pnA in 110 hours. Targets of $^{187}$Re with a thickness of 460 $\mu$g/cm$^{2}$ were sputtered on 80 $\mu$g/cm$^2$-thick carbon foils, which were facing the beam. Evaporation residues (ERs) were separated in-flight from the beam particles and implanted into a double-sided silicon strip detector (DSSD) with 48 horizontal and 128 vertical strips. The width of each strip was $\sim$1 mm, while the separation distance between them was $\sim$50 $\mu$m. Preamplifiers with fast rising time ($\sim$40 ns) were used in the experiment. Noise was reduced by cooling the preamplifiers and DSSD with (-6 to -15 $^{\textrm{o}}\textrm{C}$) alcohol and improving the grounding of the DAQ system. Details about the experimental setup can be found in Refs~\cite{Sun2016,Huang2017}.

In order to resolve short-lived products with lifetime $\leq$1 $\mu$s, a fast digital pulse processing technique was used in the experiment. Preamplified signals from each strip of DSSD were recorded in 15 $\mu$s-long traces (including nearly 2.5 $\mu$s baseline before the trigger point) with sampling frequency of 50 MHz~\cite{Sun2016}. Signals from the same strip with time difference shorter than $\sim$12 $\mu$s will pile up and are stored in a single trace. Amplitudes of pileup pulses were extracted by two algorithms according to the time differences $\bigtriangleup T$ between them, i.e., trapezoidal algorithm~\cite{Jordanov1994} for $\bigtriangleup T$ $\geq$ 0.5 $\mu$s and average difference algorithm~\cite{Sun2016} for $\bigtriangleup T$ $<$ 0.5 $\mu$s. Energy resolutions (FWHM) of all the vertical strips summed up for double/triple pileup events with time differences down to 0.5 $\mu$s and 0.2 $\mu$s were around 55 and 70 keV, respectively. For non-pileup $\alpha$-decay events, the energy resolution of DSSD was 22 keV.

\begin{figure}
\centering
  \includegraphics[width=0.35\textwidth]{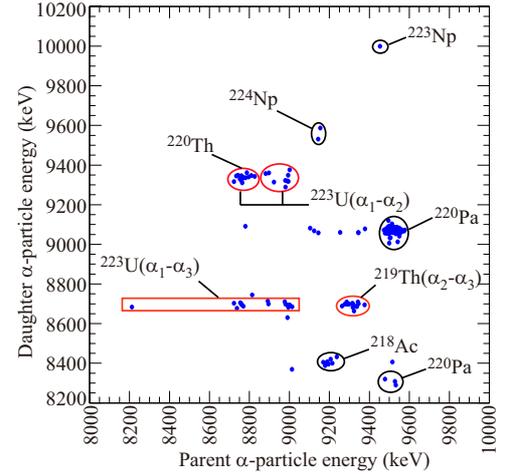}
  \caption{\label{e1e2}A two dimensional plot of $\alpha$-$\alpha$ correlations for all possible pairs of correlated chains observed in the $^{40}$Ar + $^{187}$Re reaction at a beam energy of 188 MeV. The time windows were 200 ns $\leq$ $\Delta T(ER-\alpha)$ $\leq$ 0.6 ms and 200 ns $\leq$ $\Delta T(\alpha-\alpha)$ $\leq$ 20 ms for events observed within the same DSSD pixel. The identified parent nuclei are labeled. Isotope $^{219}$Th can be produced as evaporation residue or $\alpha$-decay daughter of $^{223}$U.}
\end{figure}

Isotope $^{223}$U was unambiguously identified by establishing time-position correlated chains of $^{223}$U(ER) $\underset{\text{$18~\mu s$}}{\overset{\text{$\alpha_{1}$}}{\longrightarrow}}$ $^{219}$Th $\underset{\text{$1.05~\mu s$}}{\overset{\text{$\alpha_{2}$}}{\longrightarrow}}$ $^{215}$Ra $\underset{\text{1.66~ms}}{\overset{\text{$\alpha_{3}$}}{\longrightarrow}}$ $^{211}$Rn, where the half-lives were taken from the quoted literature~\cite{nndc,Browne2001}. Correlated implantation and $\alpha$-decay events were found in the same 1 $\times$ 1 mm$^{2}$ pixel in DSSD. Due to the short half-life of $^{219}$Th, $\alpha_1$ and $\alpha_2$ signals will pile up. The digital pulse processing of the recorded traces allows to resolve the two overlapping decay signals, to determine their energies and the time difference between them. A two-dimensional plot of $\alpha$-$\alpha$ correlations is shown in Fig.~\ref{e1e2}, which includes all possible pairs of correlated decays whether the third one exists or not. The time window between recoil and $\alpha$ decay is 200 ns $\leq$ $\Delta T(ER-\alpha)$ $\leq$ 0.6 ms and that between any two $\alpha$ decays is 200 ns $\leq$ $\Delta T(\alpha-\alpha)$ $\leq$ 20 ms, selected to highlight the identification of $^{223}$U. Another point in Fig.~\ref{e1e2} is that $\alpha$-$\alpha$ correlations of $^{220}$Th-$^{216}$Ra (produced in the $\alpha$p2n evaporation channel) resemble some of $^{223}$U in the $\alpha$-particle energies. In such case, $^{223}$U was distinguished from $^{220}$Th by the $\alpha$ decay of the granddaughter $^{215}$Ra or by the half-lives of parent ($T_{1/2}$($^{223}$U) = 18 $\mu$s and $T_{1/2}$($^{220}$Th) = 9.7 $\mu$s) and daughter ($T_{1/2}$($^{219}$Th) = 1.05 $\mu$s and $T_{1/2}$($^{216}$Ra) = 182 ns) nuclei if the third $\alpha$ particle escaped.

Besides the full-energy $\alpha$ decays of $^{223}$U and its daughter products, some partial escapes of them were also included in the analysis, as shown in the summary of measured decay chains in Table~\ref{table1}. For partially escaping decays, the corresponding energies can not be used, but their timing information is intact. Thirty-four $\alpha$-decay events were attributed to $^{223}$U. For eight of them, only partial energies of $\alpha$ particles emitted from $^{223}$U were recorded, but their daughters or granddaughters are full energy decays.

\begin{table*}
  \begin{center}
  \caption{\label{table1}Rows 2 and 3: Two $\alpha$-decay branches of $^{223}$U observed in this study. The $\alpha$-particle energy ($E_{\alpha}$), half-life ($T_{1/2}$), intensity ($I_{\alpha}$), reduced $\alpha$-decay width ($\delta^{2}$) and hindrance factor (HF) for each branch of $^{223}$U, and $\alpha$-particle energies and half-lives for the daughters $^{219}$Th and $^{215}$Ra, are listed. Our results are compared with the literature values~\cite{nndc,Browne2001} shown in the last row. The new half-life 62$^{+14}_{-10}$ $\mu$s of $^{223}$U was used for $\delta^{2}$ and HF calculations. See text for more information.}

  \begin{tabular}{cccccccccccccc}
\hline
  Branch & n$^{a)}$ & $E_{\alpha}$($^{223}$U) & $I_{\alpha}$ & $\delta^{2}$ & HF & $E_{\alpha}$($^{219}$Th) & $E_{\alpha}$($^{215}$Ra) & $T_{1/2}$($^{223}$U) & $T_{1/2}$($^{219}$Th) & $T_{1/2}$($^{215}$Ra)\\
  & & [keV]& [\%]& [keV]& & [keV]& [keV]& [$\mu$s]& [$\mu$s]& [ms]\\
  \hline

  1&17-11-9 & 8753(16) & 65(20) & 198(61) & 1.6(5) &9340(18) & 8695(4) & 50$^{+16}_{-10}$& 0.75$^{+0.24}_{-0.15}$ & 1.50$^{+0.65}_{-0.35}$\\
  2& 9-6-7 & 8993(17) & 35(13)& 25(9) & 12(5) &9336(18) & 8693(4) & 78$^{+39}_{-20}$& 1.24$^{+0.62}_{-0.31}$ & 1.60$^{+0.80}_{-0.40}$\\
  All &26-20-24 & - & - & - & - &9338(16) & 8696(3)	& 62$^{+14}_{-10}$ & 0.94$^{+0.21}_{-0.15}$ & 1.51$^{+0.40}_{-0.26}$\\

  Lit. &-&8780(40)    &100    & 178(32)& 1.1(2) & 9340(20) &8700(5)   &18$^{+10}_{-5}$           &1.05(3)             &1.66(2)\\
  \hline
  \end{tabular}

  \begin{threeparttable}
  \begin{tablenotes}
  \footnotesize
    \item[a)] Number of full energy decays for $^{223}$U - Number of full energy decays for $^{219}$Th - Number of full energy decays for $^{215}$Ra.
  \end{tablenotes}
  \end{threeparttable}

\end{center}
\end{table*}

\begin{figure}
\centering
  \includegraphics[width=0.34\textwidth]{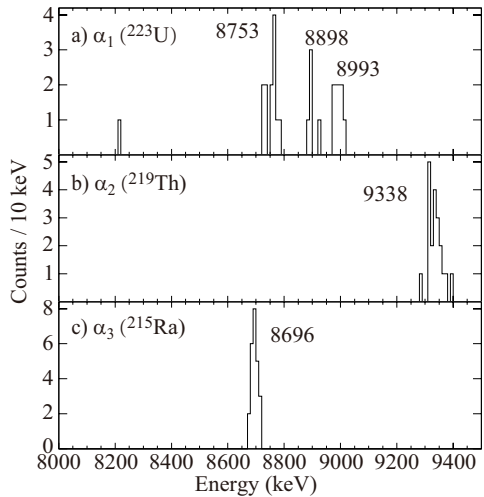}
  \caption{\label{e123}The $\alpha$ spectra of $^{223}$U, $^{219}$Th and $^{215}$Ra for correlated chains of the ER-$\alpha_{1}$-$\alpha_{2}$-$\alpha_{3}$, ER-$\alpha_{1}$-$\alpha_{3}$ or ER-$\alpha_{1}$-$\alpha_{2}$ type. All the events were found in the time windows of $\Delta T(ER-\alpha_{1})$ $\leq$ 550 $\mu$s, $\Delta T(\alpha_{1}-\alpha_{2})$ $\leq$ 11 $\mu$s and $\Delta T(\alpha_{2}-\alpha_{3})$ $\leq$ 10 ms. See text for more information.}
\end{figure}

The $\alpha$ spectra of isotopes $^{223}$U, $^{219}$Th and $^{215}$Ra identified in these $\alpha$-decay chains are shown in Fig.~\ref{e123}. Three $\alpha$ peaks at 8753(16), 8898(18) and 8993(17) keV are observed in $^{223}$U as shown in Fig.~\ref{e123} (a), with intensities of 46(15)\%, 19(9)\% and 35(13)\%, respectively, where the errors are from the statistical only. However, as will be shown below, some of these intensities will need to be corrected for the possible effect of $\alpha$ + e$^-$ summing. One 8212 keV (in vertical strip of DSSD) $\alpha$-decay event of $^{223}$U was attributed to a crosstalk, because a 8602 keV energy produced by the same event in horizontal strip was recorded. Moreover, based on GEANT4 Monte Carlo simulations, the 8212 keV event is very unlikely to result from partially escaping 8753, 8898 and 8993 keV $\alpha$ particles, as in the region of (8212 $\pm$ 70) keV only 0.03 event is expected corresponding to the 26 full energy events of these three peaks. The $\alpha$-particle energies for $^{219}$Th and $^{215}$Ra were determined to be 9338(16) and 8696(3) keV, respectively, consistent with literature values.

\begin{figure*}
\centering
  \includegraphics[width=0.82\textwidth]{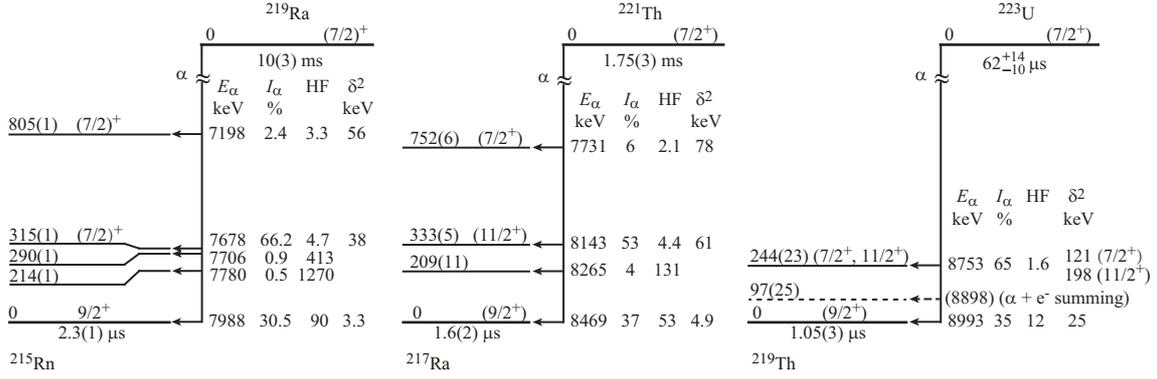}
  \caption{\label{levels}The $\alpha$-decay schemes of $N$ = 131 isotones $^{219}$Ra~\cite{Browne2001}, $^{221}$Th~\cite{Kondev2018} and $^{223}$U deduced in this study. The 244(23) keV excited state observed in the $\alpha$ decay of $^{223}$U was proposed to be the counterpart of the ($7/2^{+}$) and ($11/2^{+}$) excited states in $^{215}$Rn (and $^{217}$Ra). The 97(25) keV level is drawn with dashed line because the corresponding 8898 keV decay may be produced by $\alpha$(8753 keV) + e$^-$ summing. See text for details.} 
\end{figure*}

Half-lives for the 8753, 8898 and 8993 keV $\alpha$ peaks of $^{223}$U were extracted to be 60$^{+24}_{-13}$, 28$^{+23}_{-9}$ and 78$^{+39}_{-20}$ $\mu$s, respectively, by taking into account types of $\alpha_1$-$\alpha_2$-$\alpha_3$, $\alpha_1$-$\alpha_2$, $\alpha_1$-$\alpha_3$, and $\alpha_2$-$\alpha_3$ correlations. Due to their very low statistics, the errors of half-lives were calculated using the maximum likelihood method described in Ref.~\cite{Schmidt1984}. These half-lives are consistent within the error bars. So all these three $\alpha$-decay groups are assigned as originated from the ground-state decay of $^{223}$U, feeding different levels of $^{219}$Th. Combining all the thirty-four events together, the half-life of $^{223}$U was determined to be 62$^{+14}_{-10}$ $\mu$s, larger than the previous measurement~\cite{Andreyev1991}. The decay scheme of $^{223}$U is shown in Fig.~\ref{levels}. To determine the excitation energies of the populated levels in $^{219}$Th, $Q_{\alpha}$ values of $^{223}$U were calculated.

In the previous measurements of $^{223}$U using analog electronics~\cite{Andreyev1991}, a single $\alpha$-decay branch at 8780 keV was reported based only on the full-energy summing signals from $^{223}$U + $^{219}$Th decay. This $\alpha$-particle energy seemed to be a ground-state to ground-state transition as the evaluation in the latest A = 223 Nuclear Data Sheets~\cite{Browne2001}. Our new results suggest that the 8780 keV value most likely corresponds to the strongest 8753 keV ground-state to 244 keV excited state transition in this work.

A comment on the 8898 keV decay, which would establish an excited state at 97 keV in $^{219}$Th, should be presented here. We cannot exclude that this peak is due to the summing of the 8753 keV $\alpha$ particle with the subsequent conversion electron, originated from internal conversion of the 244 keV $\gamma$-ray transition. Indeed, if $\alpha$ decay feeds a low-lying excited state, which de-excites by a strongly converted $\gamma$-ray transition, a well-known effect of energy summing of $\alpha$ decay and subsequent conversion electron can happen, if both $\alpha$ particle and conversion electron are measured in the DSSD~\cite{Hessberger1989,Andreyev2004}. This would lead to the appearance of `artificial' summing $\alpha$ lines in the spectra. The total conversion coefficients of the 244 keV transition are: $a_{tot}$(M1) = 1.5, $a_{tot}$(E1) = 0.06 and $a_{tot}$(E2) = 0.33~\cite{anu}. Therefore, in case of E1 or E2 multipolarity, the conversion of the 244 keV transition is negligible, thus no $\alpha$(8753 keV) + e$^-$ summing can be observed, and the 8898 keV will be a real $\alpha$ peak. For M1 the K-conversion is dominant ($a_K$(M1) = 1.19), resulting in K-electron energy of 134 keV. The $\alpha$(8753 keV) + e$^-$(134 keV) summing would indeed give rise to a summing peak around 8887 keV, close to the observed peak at 8898 keV. 

\begin{figure}
\centering
  \includegraphics[width=0.47\textwidth]{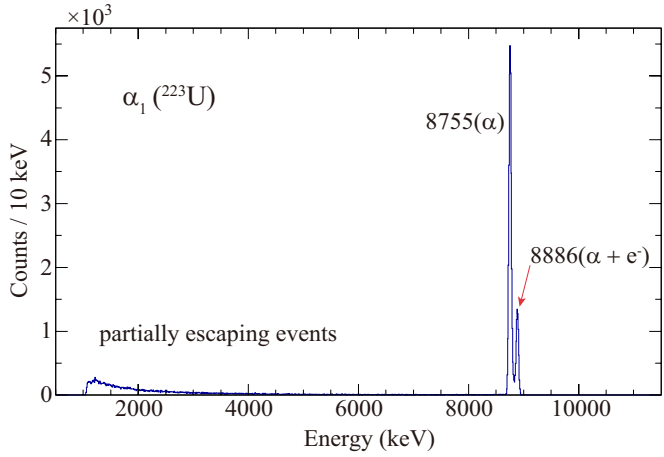}
  \caption{\label{summing}The Geant4 Monte Carlo simulation for $\alpha$(8753 keV) + e$^-$(134 keV) summing in DSSD. The pure $\alpha$ peak and $\alpha$ + e$^-$ peak are labeled with the fitting results. See text for more information.}
\end{figure}

Based on the GEANT4 Monte Carlo simulations for $\alpha$ + e$^-$ summing in DSSD (see Fig.~\ref{summing}), we estimated that the whole $\sim$8898 keV decay can be explained as an artificial peak being due to $\alpha$(8753 keV) + e$^-$ summing of the 134 keV conversion electron from the 244 keV transition if it is of pure M1 multipolarity. In the simulation, ratio between source intensities of electron and $\alpha$ particle was set to be 0.54, defined by the conversion coefficient $a_K$(M1) = 1.19. A relative intensity of 26\% between $\alpha$ + e$^-$ peak and pure $\alpha$ peak was obtained, which is nearly consistent with the experimental result 41(24)\%. So there seems no level at 97 keV in $^{219}$Th and the intensity of the 8753 keV $\alpha$ decay should be 65(20)\%\footnote{The earlier extracted half-life of $^{223}$U should not be affected by the conversion-electron process, which is assumed to be comparatively short.}, where the reported error is statistical only, see Table~\ref{table1}. We will return to this issue in the Discussion section. Moreover, since the known lowest excited states of $N$ = 129 isotones $^{215}$Rn and $^{217}$Ra are above 200 keV, the 97 keV excited state of $^{223}$U seems too low. 

\section{\label{sec:level3}Discussion}

To understand the observed fine structure decays of $^{223}$U, we also show the $\alpha$-decay schemes of $N$ = 131 isotones $^{219}$Ra and $^{221}$Th in Fig.~\ref{levels}. These two isotones are located at the edge of the region where octupole deformation was predicted~\cite{Moller2016}. A 7/2$^+$ ground state was reproduced for each of $^{219}$Ra and $^{221}$Th by reflection-asymmetric rotor model~\cite{Leander1988} at $\beta_2$ = $\beta_3$ = 0.1 deformation. 

Spins and parities have been tentatively assigned to the populated excited states in $^{215}$Rn~\cite{Browne2001} and $^{217}$Ra~\cite{Kondev2018}. Even though (7/2)$^+$ and (11/2$^+$) are assigned to the 315 keV level in $^{215}$Rn and 333 keV level in $^{217}$Ra, respectively, both spins are possible for each of them. The possible spins of the 315 keV state in $^{215}$Rn are (7/2, 11/2)$^+$ according to the measured $\gamma$ transition mode~\cite{El-Lawindyt1987,Hackett1989}, the same applies for the 333 keV state in $^{217}$Ra. Decays to these two levels are weakly hindered with HF $<$ 5, implying similar configuration between initial and final states. The (7/2$^+$) and (11/2$^+$) assignments were interpreted to be of octupole configuration in the reflection-asymmetric rotor model, while in the shell model framework they were supposed to be originated from the same configuration $\nu g_{9/2}^{2}i_{11/2}$~\cite{Sheline1994}. Branches feeding the spherical 9/2$^+$ ground states of $^{215}$Rn and $^{217}$Ra were strongly hindered with HF = 90 and 53, respectively, echoing the non-spherical octupole configuration of the ground states in $^{219}$Ra and $^{221}$Th. 

In the $\alpha$ decay of $^{223}$U, the 8753 keV transition feeding the 244 keV (`apparent intensity' $I_{\alpha}$ = 46\%) level in $^{219}$Th is unhindered and shows hindrance factor of 2.2. In case the 8898 keV peak is due to $\alpha$ + e$^-$ summing, as proposed in the previous section, the intensity of 8753 keV decay increases to 65\%, leading to HF(8753 keV) = 1.6. We use this intensity value in Table~\ref{table1} and Fig.~\ref{levels}. Thus in either scenario (with or without $\alpha$ + e$^-$ summing), the 8753 keV decay is unhindered. The 8993 keV ground-state to ground-state transition of $^{223}$U is hindered with HF = 12. Based on these HF values, spin of the ground state of $^{223}$U should be the same as that of the 244 keV excited state of $^{219}$Th, while is different from the (9/2$^+$) ground state of $^{219}$Th. Based on the intensities and hindrance factors, (7/2$^+$) and (7/2$^+$, 11/2$^+$) are tentatively assigned to the ground state of $^{223}$U and the 244 keV excited state of $^{219}$Th by following the systematics, respectively. These assignments are also beneficial to the discussion of the evolution of systematics. 

In the in-beam $\gamma$ spectroscopy of $^{219}$Th~\cite{Reviol2009}, a 362 keV level with a tentative assignment of J$^{\pi}$ = (11/2$^+$) was thought to correspond to the (11/2$^+$) level at 333 keV in $^{217}$Ra. This level was observed in both $\alpha$ decay and in-beam $\gamma$ spectroscopy, while the 362 keV level in $^{219}$Th was observed only in the in-beam $\gamma$ spectroscopy. Because fusion-evaporation reactions like that in Ref.~\cite{Reviol2009} are tend to populate near-yrast levels whereas the $\alpha$-decay don't, it is possible both the 362 and 244 keV levels exist in $^{219}$Th. Another possible explanation is the onset of structure change in the ground state of $^{223}$U.

\begin{figure}
\centering
  \includegraphics[width=0.33\textwidth]{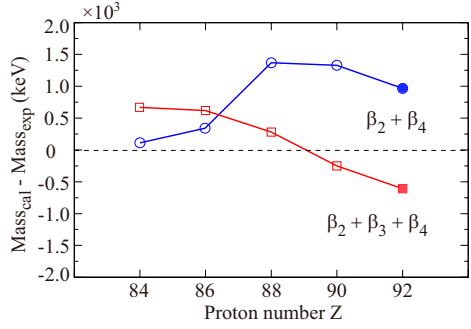}
  \caption{\label{mass}Differences between calculated and experimental masses for even-$Z$ isotones with $N$ = 131. The experimental masses are deduced from the ground-state to ground-state $Q_{\alpha}$ values. A comparison is made between calculated masses with (red squares) and without (blue circles) including octupole correlations $\beta_{3}$. In both cases, the symmetric shape coordinates $\beta_{2}$ and $\beta_{4}$ were considered.}
\end{figure}

The mass of $^{223}$U was deduced from the ground-state to ground-state $Q_{\alpha}$ obtained in the present work. In Fig.~\ref{mass}, the experimental masses of $N$ = 131 isotones up to $^{223}$U were compared with two versions of theoretical calculations, one taking into account the octupole correlations~\cite{Moller2016} while another not~\cite{Moller1981}. For $^{219}$Ra($Z$ = 88) and heavier isotones, the theoretical masses with octupole correlations around $\beta_3$ = -0.134~\cite{Moller2016} become favorable and reproduce the experimental data better, suggesting the onset of octupole deformation at $Z$ = 88. Until to $Z$ = 92, both models do equally poor job for $^{223}$U, the weakening of octupole deformation is implied.

\begin{figure}
\centering
  \includegraphics[width=0.33\textwidth]{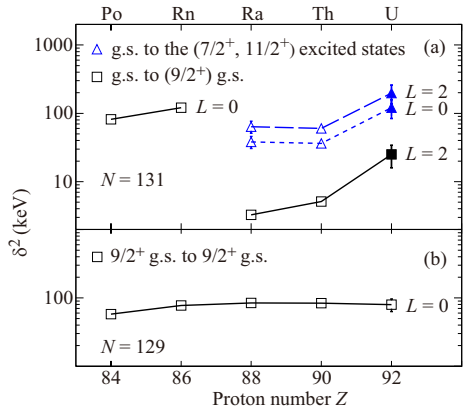}
  \caption{\label{delta2}Reduced $\alpha$-decay widths ($\delta^{2}$) for the $N$ = 131 (a) and 129 (b) isotones with even $Z$ between 84 and 92. The $\delta^{2}$ values extracted from the new $\alpha$-decay data of $^{223}$U are shown with full symbols, the errors are mainly from the relative intensities.}
\end{figure}

To understand the evolution of nuclear structure, reduced $\alpha$-decay width ($\delta^{2}$) is used. It can probe the overlap between the wave functions of the initial and final states of $\alpha$ decay. It is calculated using the relation $\lambda$ = $\delta^{2}P/h$ \cite{Rasmussen1959,Rasmussenodd1959}, where $\lambda$ is the measured partial decay constant, $h$ is the Planck's constant and $P$ is the penetration probability of $\alpha$ particle with angular momentum $L$ through the Coulomb and centrifugal barriers. The reduced $\alpha$-decay widths for $N$ = 131 and 129 isotones are shown Fig.~\ref{delta2}. No obvious fine structure are observed in the $\alpha$ decays of the $N$ = 129 isotones and their reduced widths remain nearly constant around 80 keV, reflecting similar spherical $\nu g_{9/2}$ configurations in the ground states of $N$ = 129 parents and $N$ = 127 daughters.

For the first two $N$ = 131 isotones, $^{215}$Po and $^{217}$Rn (see Fig.~\ref{delta2} (a)), the ground-state to ground-state decays proceed between the $\nu g_{9/2}$ configurations with $\delta^{2}$ values of $\sim$100 keV, in agreement with the very similar spherical structures between parent and daughter nuclei. From $Z$ = 88 on, fine structure with competing branches appear and the reduced widths for the ground-state to ground-state transitions drop suddenly. Decays of $^{219}$Ra and $^{221}$Th show strong hindrance with $\delta^{2}$ $\sim$ 5 keV, indicating very different structure between the ground states of parent and daughter nuclei. New $\delta^{2}$ value for the ground-state to ground-state transition of $^{223}$U continues the rising trend in $^{219}$Ra and $^{221}$Th, but still shows hindrance with $\delta^{2}$ = 25 keV. If other assignment rather than (7/2$^+$) to the ground state of $^{223}$U is used, the rising trend can not be changed. A reverse movement is revealed in the hindrance factors.

The rising trend of reduced $\alpha$-decay width for the ground-state transitions of $N$ = 131 isotones from $Z$ = 88 to 92, combining with the behavior of nuclear masses in Fig.~\ref{mass}, may indicate a weakening octupole deformation in the ground state of $^{223}$U.

The $\delta^{2}$ values for the most dominant branch with possible assignments of (7/2$^{+}$,11/2$^{+}$) are shown in Fig.~\ref{delta2} (a), the relative large values imply that octupole correlations are developing in the excited levels of the $N$ = 129 daughter nuclei. This is in agreement with the evidences of octupole deformation in the in-beam $\gamma$-spectroscopy studies of $^{217}$Ra~\cite{Sugawara1987,Sugawara1984} and $^{219}$Th~\cite{Reviol2009}.

\section{\label{sec:level4}Summary and outlook}
Two $\alpha$-decay branches of $^{223}$U feeding the ground state and 244 keV excited state of $^{219}$Th are reported. An $\alpha$-particle energy of 8993 keV and a half-life of 62$^{+14}_{-10}$ $\mu$s are assigned to the ground-state decay of $^{223}$U. Evolution of nuclear structure for $N$ = 131 isotones is discussed in terms of nuclear mass and reduced $\alpha$-decay width, structure in the ground state of $^{223}$U may go through a weakening octupole deformation relative to its lighter isotones $^{219}$Ra and $^{221}$Th.

\section{Acknowledgements}
The authors would like to thank the accelerator crew of HIRFL for providing the stable $^{40}$Ar$^{7+}$ beam of high intensity. This work was supported by the National Natural Science Foundation of China (Grant Nos. 11635003, 11675225, U1632144, 11405224, 11435014, 11505035 and 11375017), the National Key R\&D Program of China (Contract No. 2018YFA0404400), the Hundred Talented Project of the Chinese Academy of Sciences and the Key Research Program of the Chinese Academy of Sciences (Grant No. XDPB09), Science and Technology Facilities Council (STFC) of the United Kingdom.

\end{document}